\documentclass[aps,preprint,amsfonts,preprintnumbers,superscriptaddress,showpacs,nofootinbib]{revtex4}
\usepackage{graphics}
\usepackage{amsmath}
\usepackage{amssymb}
\usepackage{color}
\usepackage{bm}
\usepackage{hyperref}

\begin{document}
\title{Misconceptions on Effective Field Theories and spontaneous symmetry breaking: Response to Ellis' article}
\author{Thomas Luu}
\affiliation{Institute for Advanced Simulation (IAS-4), and J\"ulich Center for Hadron Physics,
  Forschungszentrum J\"ulich, Germany}
\affiliation{Helmholtz-Institut f\"ur Strahlen- und Kernphysik and Bethe Center for Theoretical Physics,
  Rheinische Friedrich-Williams-Universit\"at Bonn, Germany}
\author{Ulf-G. Mei\ss ner}
\affiliation{Helmholtz-Institut f\"ur Strahlen- und Kernphysik and Bethe Center for Theoretical Physics,
  Rheinische Friedrich-Williams-Universit\"at Bonn, Germany}
\affiliation{Center for Science and Thought, Rheinische Friedrich-Williams-Universit\"at Bonn, Germany}
\affiliation{Institute for Advanced Simulation (IAS-4), and J\"ulich Center for Hadron Physics,
  Forschungszentrum J\"ulich, Germany}
\affiliation{Tbilisi State University, 0186 Tbilisi, Georgia}

\date{\today}

\begin{abstract}
In an earlier paper~\cite{Luu:2019jmb} we discussed emergence from the context of effective field theories,
particularly as related to the fields of particle and nuclear physics.  We argued on the side of reductionism
and weak emergence. George Ellis has critiqued our exposition in~\cite{Ellis:2020vij}, and here we provide our
response to his critiques.  Many of his critiques are based on incorrect assumptions related to the formalism
of effective field theories and we attempt to correct these issues here.  We also comment on other statements
made in his paper. Important to note is that our response is to his critiques made in archive versions
arXiv:2004.13591v1-5 [physics.hist-ph].  That is, versions 1-5 of this archive post.  Version 6 has
similar content as versions 1-5, but versions 7-9 are seemingly a different paper altogether
(even with a different title).
\end{abstract}

\maketitle


\section{Introduction}
Emergent phenomena are fascinating quantities that when viewed from lower level (more fundamental) constituents
are highly complex in nature.  The question of whether such phenomena can be totally derived from their underlying
constituents provides an intriguing debate not only from a  philosophical point of view, but also from a
scientific point of view that spans all areas of science.  At issue, from our point of view, is the invocation
of \emph{strong} emergence, which is the notion that certain (or all) properties of some emergent phenomena
cannot be deduced from its fundamental constituents.  The weaker claim, naturally called \emph{weak} emergence,
states that all properties of an emergent phenomena can, \emph{in principle}, be deducible from its fundamental
constituents~\footnote{Though weak emergent phenomena can \emph{in principle} be deduced from fundamental
  constituents, such a deduction may \emph{in practice} not be possible.}.  A more detailed definition of
these cases is given in~\cite{Chalmers2006-CHASAW}.

In our paper~\cite{Luu:2019jmb} we gave an overview of the formalism of Effective Field Theories (EFT) and
its relation to emergent phenomena, providing examples prevalent in nuclear and particle physics.  We won't
repeat our arguments here, but only state that in our paper we rejected the concept of strong emergence and
argued in favor of weak emergence and reductionism in this context.  Ellis has since criticised our
arguments in~\cite{Ellis:2020vij}, and we take this opportunity to respond to his criticisms.

Ellis' criticisms have not changed our views on the matter.  Furthermore, many of his criticisms are
based on incorrect assumptions and an incomplete understanding of the EFT formalism.  Thus our response
here also serves as a means to clarify any misconceptions that might occur from an initial read of Ellis'
critiques by anyone uninitiated in the concepts of EFTs.  

Our paper is simply structured.  In the next section~\ref{sect:responses} we respond to Ellis' critiques,
pointing out incorrect statements he makes related to EFTs that lead to wrong conclusions.   We make
some further comments in section~\ref{sect:comments} that are not directly related to Ellis' statements related to
EFTs, but we feel are important nonetheless in view of the debate on strong emergence.  We conclude
in section~\ref{sect:conclusions}. 

\section{Our Response\label{sect:responses}}

\subsection{On the connection between an EFT and its lower-level fundamental theory}
Ellis states that EFTs \emph{ ``. . . are an approximation and are not strictly derived from a more
  fundamental theory."}  He goes on to question whether EFTs, because of this limitation, can be used
to even defend reductionism.  To make his point he pulls quotes from various sources (Ref.~\cite{Hartmann:2001zz}
in particular) to demonstrate that EFTs often are no better than some ad hoc model of the system in question.  

{\bf This is categorically wrong.}  Effective field theories, defined as we outlined in~\cite{Luu:2019jmb},
are often derived from a fundamental theory (we simplify the discussion for the moment, as cases exist, where this
fundamental theory is not (yet) known and even might not be needed).  A prime example is non-relativistic
Quantum Electrodynamics (NRQED), which is an effective field theory of bound states, i.e.
emergent phenomena, in Quantum Electrodynamics (QED)~\cite{Caswell:1985ui}.  Here the coefficients of the
low-energy effective field theory can be directly matched to non-pertubative resummations (i.e. calculations) of the
fundamental theory, which is QED.  NRQED and QED share the same symmetries, but what differentiates NRQED
from QED is the reshuffling of diagrammatic terms such that operators representing the bound state degrees
of freedom are explicit in NRQED.  Such a reshuffling of terms results in low-energy constants (LECs) with
each associated operator in NRQED.   To calculate emergent bound states with fundamental QED requires
an infinite number of resummations.  Not surprisingly, the number of operators in NRQED is also infinite,
but with a defined separation of scales there is a systematic hierarchy in relevance of these
operators.  One then truncates the number of terms in the EFT to achieve a desired accuracy.
The truncation of terms to achieve a desired accuracy can be viewed as an ``approximation"  of the
fundamental theory, but it is \emph{in principle} possible to perform calculations with the EFT to any
desired accuracy.  Note also that in the fundamental theory there are practical limitations in performing
calculations beyond some accuracy, as best exemplified in the tenth order calculation of the electrons
anomalous magnetic moment~\cite{Aoyama:2017uqe}, but this is another story.

Coming back to nuclear and particle physics, we admit that an explicit derivation of chiral
perturbation theory ($\chi$EFT), the low-energy EFT of Quantum Chromodynamics (QCD) which governs quarks
and gluons, from QCD is difficult to do, partly because the degrees of freedom of the emergent phenomena
(pions and nucleons) are vastly different from their fundamental constituents. However, we stress that
it has been shown that the Greens functions of QCD are indeed exactly reproduced by the ones in
chiral perturbation theory~\cite{Leutwyler:1993iq}, which leaves us with the LECs, as
different theories can in principle
lead to the same operator-structure but are characterized by different LECs. In QCD,
the LECs of the EFT are very difficult to extract formally, but can, for example, be determined
from non-perturbative numerical methods.  Equally valid is the determination of the coefficients
from empirical data, which is what is commonly done.  It is quite amazing that these days some
of the LECs can be determined more precisely from lattice QCD calculations than from phenomenology, which
is related to the fact that one the lattice we can vary the quark masses but not in Nature.
Either way, the principles are the same:  The
operators in the EFT share the same symmetry as the fundamental theory, and the separation of scales
dictates the rate of convergence of the terms and thus how many terms are required for a desired accuracy.
The EFT \emph{is derived} from the underlying theory, and \emph{in principle} can calculate observables to
arbitrary precision. Clearly, scale separation is an important ingredient in any EFT. This will
be stressed at various points in what follows. Another important remark is that chiral perturbation
theory is what is called a {\em non-decoupling EFT}, as the relevant degrees of freedom, the
Goldstone bosons, are only generated through the spontaneous symmetry breaking as discussed below.

The notion that EFTs are glorified models with multiple fit parameters (LECs) has been a misconception
since its first application in nuclear physics by Weinberg~\cite{Weinberg:1991um}. Another historical
example is the abandonment of the Fermi theory of the weak interactions because it violates
unitarity at about 80~GeV (at that time a dream for any accelerator physicist). Now we know
that it is indeed an EFT, with the breakdwon scale given by the mass of the heavy vector
bosons, alas 80~GeV. As stated above, the
LECs of the theory are not {\em ad hoc} (the same cannot be said of many models in other areas of
physics!) but are directly connected to the underlying theory.  Furthermore, the number of LECs per
given order of calculation is predetermined, and because of its strict power counting rules,
estimates of the theory's \emph{error} due to omission of higher order terms can be made.
The same cannot be said for any type of model.  The survey by Hartmann~\cite{Hartmann:2001zz}
completely overlooks these facts and thus gives a very poor and inaccurate description of the
efficacy of EFTs.  For a survey with philosophical connections, the reader should consider~\cite{Rivat:2020amd}.
Note also that EFTs are widely used in all field of physics, in atomic, cold atom, condensed matter,
astrophysics, you name it. This shows that EFTs are of general interest as they capture the pertinent
physics in a certain energy regime. This, however, does not mean that they are all disconnected,
often the reduction in energy and thus resolution leads to a tower of EFTs that fulfill matching
conditions in the course of the reduction in resolution. A nice example is flavor-diagonal CP violation,
where one starts with a beyond the standard model theory, like e.g. supersymmetry, at scales way above
the electroweak breaking and runs down through a series of EFTs to the chiral Lagrangian of pions
and nucleons including CP-violating operators, see e.g.~\cite{Dekens:2014jka}. Note that while the
scale of supersymmetry is about 1~TeV, the one of the chiral Lagrangian is well below 1~GeV, thus
one bridges about 3 orders of magnitude by this succession of EFTs.

\subsection{On the applicability of EFTs to other areas of science}
Ellis asks if EFTs can be applied \emph{``. . . to quantum chemistry, where methods such as
  the Born-Oppenheimer approximation and Density functional Theory (DFT) have been used?"}  

We see no reason why EFTs cannot be applied here.  Indeed, both of these methods are mean-field
approximations which, from an EFT perspective, are the leading order term of some
EFT~\cite{Furnstahl:2019lue,Brambilla:2017uyf}.  In this view, an EFT description goes \emph{beyond}
both DFT and Born-Oppenheimer approximations, since it naturally includes dynamics of quasi-particles
(emergent phenomena) above the mean-field approximation. To be more specific, the Born-Oppenheimer approximation
shares all the features of an EFT, the light (fast) modes are decoupled form the heavy (slow) ones,
all symmetries pertinent to the interactions are included and the proper degrees of freedom are
identified. What is simply missing is to set up a power counting in which to calculate the corrections.
This has recently been achieved in hadron physics, where the Born-Oppenheimer approximation has been in use for quite some time, but has since been surpassed by EFTs. For example, it was heavily utilized in the context of the bag model~\cite{Hasenfratz:1977dt}
leading to various model studies like e.g. by one of the authors~\cite{Meissner:1986dn}. More recently,
in the context of heavy quark physics, this was even cast in terms of an EFT~\cite{Brambilla:2017uyf},
which clearly shows that even in quantum chemistry the formulation of an EFT embodying the Born-Oppenheimer
approximation should be possible. We point out that in quantum chemistry powerful techniques exist to
perform extremely precise calculations like the already mentioned DFT approaches or the coupled-cluster
scheme, originally invented for nuclear physics, thus the need for setting up an EFT has been less
urgent than in strong interaction physics. However, as DFT in chemistry now also enters the stage to
accomodate strong electronic correlations {\em ab initio}, this will change in the future.  We will return to the topic of the Born-Oppenheimer approximation when we discuss phonons in the next section.

Ellis goes on to ask about the applicability of EFTs to signal propagation in neurons,
neural networks~\footnote{Already statistical field theoretical techniques are being applied to neural
networks (see for example~\cite{buice_2013,Helias:2019pmr}).  Such theories are amenable to EFTs.
}, Darwinian evolution, and election results.   We admit we cannot
answer these questions because some of these systems fall well outside our purview of expertise. We say ``some''
because one of us, however, has embarked in research in modelling neuron dynamics~\cite{Stapmanns:2020vwg},
and in fact we are presently setting up a simulation laboratory at Forschungszentrum J\"ulich that
deals with the application of numerical quantum field theory to complex systems in particle and nuclear
physics, solid-state physics, and also biological systems like the brain.
Nevertheless, for certain fields, such numerical methods are not yet available or only based
on simple modelling, but we do not dismiss the possibility of applicability just because of our
ignorance.  Ellis, on the other hand, answers with a definite \emph{NO!} because he claims these are strong emergent phenomena. 

\subsection{Spontaneous Symmetry Breaking and Topological Effects in EFTs}
Arguably the most egregious error that Ellis makes, from our point of view, is the statement that EFTs
cannot capture spontaneous symmetry breaking (SSB), or explicit broken symmetries in general,
and therefore cannot describe the emergent phenomena (which he claims are strong emergent) that ensue
from these reduced symmetries.  He uses the solid-state example of the reduced point symmetry of a
lattice that ultimately leads to the creation of phonons.  Indeed, he claims most of condensed matter and
solid-state physics is off-limits to EFTs, because much of their emergent phenomena is due to SSB
or explicit symmetry breaking.

We point out, however, that spontaneous symmetry breaking, and its ensuing consequences, is not 
solely relegated to the fields of condensed matter and solid-state physics, despite being the origin of
its development~\cite{Nambu:1960tm}.  One of the most prominent examples of SSB comes from the non-zero 
vacuum expectation value (VEV) of the scalar condensate in the QCD vacuum.  While the Hamilton  operator
of the theory has the chiral symmetry  SU$(N_f)_L \times$SU($N_f)_R$, in  the QCD vacuum this symmetry is
``hidden", that is,  the \emph{global} symmetry of SU$(N_f)_L\times\operatorname{SU}(N_f)_R$ is spontaneously
broken to SU$(N_f)_V$~\footnote{The interested reader might consult
e.g.~\cite{Cheng:1985bj,Donoghue:1992dd,Weinberg:1996kr} for a
detailed description of this process and the meaning of these symmetries.}.
As Ellis correctly points out, this symmetry breaking leads to emergent phenomena, which are gapless long-distance
(pseudo-)scalar modes in the theory with reduced symmetry.  The number of gapless, or massless, states is given
by the much celebrated Goldstone theorem~\cite{Goldstone:1961eq}, which states that for each generator of the
hidden symmetry that is broken there corresponds a long-distance massless (pseudo-)scalar particle.  These states
are more commonly known as Nambu-Goldstone bosons.  In QCD these bosons are the three (nearly) massless pions
(for $N_f=2$) observed in Nature.  The basis of chiral perturbation theory is exactly this theorem -
it provides the EFT of light-quark QCD with its
proper degrees of freedom (for the heavy quark sector, a different type of EFT comes into play, which we will 
not discuss here). The emergent phenomena of pions due to SSB (and coupled eventually to nucleons, which act
as matter fields) are the explicit degrees of freedom in this theory.  Thus SSB and its subsequent reduced
symmetry are not only \emph{captured} but exactly reproduced (as stated above) in this EFT, as well as any other unaffected symmetries.

The same happens when, during the formation of a crystal, whether natural or synthetic, the underlying global
Lorentz and Galilean symmetries are spontaneously broken to some reduced point symmetry of the crystal lattice.
There are some differences between relativistic and non-relativistic formulations of the Goldstone
theorem~\footnote{See e.g. the lucid discussion in Ref.~\cite{Leutwyler:1993gf}}, but the central point remains
the same: The broken generators of the underlying continuous symmetries result in the formation of (pseudo-)scalar
Nambu-Goldstone bosons, in this case the so-called phonons.  Furthermore, Goldstone's theorem, and the
subsequent phonon EFTs  based off this theorem, go on to describe the interactions \emph{between}
phonons~\cite{Leutwyler:1996er,SCHAKEL2011193}.  One can imagine adding electron degrees of freedom as matter fields in such an EFT,
all the while ensuring that the electrons respect the relevant point symmetry of the lattice, analogous to
what is done with nucleons in $\chi$EFT.  The LECs corresponding to electron-phonon interactions in the EFT can
be constrained empirically or, \emph{in principle}, calculated directly from QED with appropriate boundary conditions.  This EFT captures, at its lowest orders, the Born-Oppenheimer approximation for electrons.   Once the LECs are determined,
the phonons themselves can be integrated out of the theory, which results in an effective electron-electron
\emph{attractive} interaction~\cite{Polchinski:1992ed}~\footnote{The analog of this process in chiral perturbation theory $\chi$EFT,
  that is integrating out  pions, results in pionless EFT, see e.g.~\cite{Bedaque:2002mn}.}.

With regards to Ref.~\cite{Polchinski:1992ed} written by Polchinski in 1992, Ellis quotes various passages from this reference that seemingly lend credence to his claim that EFTs cannot be applied to high-$T_c$ superconductors.  We point out that this article was written as a series of introductory lectures on EFTs and the renormalization group.  The underlying premise of these lectures was that the electrons were approximated as a fermi liquid.  Non-fermi liquid behavior, which most certainly underlies unconventional superconductors like high-$T_c$ superconductors, is thus not captured by these EFTs.  From an EFT point of view, the separation of scale between the Debye length and inverse fermi momentum becomes less clear as electrons become more strongly correlated in non-fermi liquids.  This signifies that the degrees of freedom of the EFTs in Polchinski's lectures are, at the very least, incomplete.  Indeed, Polchinski points this out, stating the possibility that other degrees of freedom like anyons, or spin fluctuations, are required in a successful EFT of high-$T_c$ superconductors.  Thus Ellis' chosen quotes, when taken out of context, belie this point. Though there still does not exist a successful EFT that captures high-$T_c$ superconductors, there has been progress in formulating EFTs for finite-density systems~\cite{Hammer:2000xg} with quasi-particle excitations~\cite{Platter:2002yr,Kapustin:2018dch}, like magnons~\cite{Brugger:2006dz}, as additional degrees of freedom.  

Ellis also states that topological effects, prevalent in low-dimensional condensed matter and solid-state systems
such as topological insulators and superconductors, cannot be captured by EFTs because they cannot be derived from a ``bottom-up" framework.  Again this is incorrect.
One need only consider the decay process $\pi^0\to\gamma\gamma$  and understand how this process is described
in an EFT.  On classical grounds, the decay of the neutral pion to two photons is not allowed, but quantum
fluctuations allow such a decay through what is called the axial anomaly~\cite{Bell:1969ts}~\footnote{A
satisfactory description of this anomaly, and its ramifications goes beyond the point of this article, but a
nice overview is given in~\cite{Bertlmann:1996xk} for the interested reader. }.  This process is non-local and
is captured in an EFT by the Wess-Zumino-Witten (WZW) effective action, first presented in a
top-down manner by Wess and Zumino in~\cite{Wess:1971yu} and later derived by Witten~\cite{Witten:1983tw} in a \emph{bottom-up} fashion~\cite{,Hill:2009wp}.
This action is topological in nature.  Its geometrical interpretation relies on the fact that the physics
it describes is confined to the surface (where our world ``resides") of a five-sphere $S^5$. 
The WZW formalism for anomaly physics is universal and thus not constrained to the realm of
particle and nuclear physics.  Indeed, its lower-dimensional analogs are utilized in the classification
of topological insulators and superconductors~\cite{Ryu:2010zza} which Ellis claims is beyond the
purview of EFTs.  Its application naturally leads to the Chern-Simons effective Lagrangian~\cite{Harvey:2007ca} that accounts
for the integer quantum hall effect~\cite{Qi:2008ew}. 

Another area of interplay of spontaneous symmetry breaking, topology (here the one of the QCD vacuum)  and EFT is axion physics.
The QCD vacuum exhibits a non-trivial topology, with different sectors given in terms of an integer
winding number. Instanton effects allow for transitions between these sectors, which are, however,
exponentially suppressed~\cite{tHooft:1976snw}. Another ramification is the appearance of the so-called
$\theta$ term in QCD, which leads to CP violation. However, the value of the $\theta$ parameter accompanying this
term is smaller than $10^{-11}$ as deduced e.g. from the experimental upper limit on the neutron
electric dipole moment~\cite{Abel:2020gbr}. This constitutes the so-called ``strong CP problem''.
An elegant solution was proposed by  Peccei and Quinn~\cite{Peccei:1977hh}, who elevated this parameter
to a dynamical variable related to a new U(1) symmetry, nowadays called U(1)$_{\rm PQ}$. This symmetry is
spontaneously broken with the appearance of a Goldstone-like particle, the axion, whose interactions with light and matter can be cast into an EFT. For a recent high-precision calculation of the axion-nucleon coupling, see~\cite{Vonk:2020zfh} which also contains references to earlier work in this important area of beyond-the-standard-model (BSM) physics, that has become one of the major playgrounds for EFTs.

Thus, EFTs can indeed capture both SSB and topological effects, and therefore their subsequent emergent phenomena.

\section{Additional comments\label{sect:comments}}

\subsection{Spontaneous Symmetry Breaking at the micro versus the macro scale}
Ellis differentiates SSB that occurs at the \emph{micro} scale and SSB that occurs at the \emph{macro} scale.
He labels them  SSB(m) and SSB(M), respectively.  As examples, he mentions the Higgs mechanism as a source
for SSB(m) while the SSB that leads to crystallization is an example of SSB(M).  Presumably the fact
that the crystal exhibits long-range order at the macro scale is the reason why it falls under SSB(M).   

Ellis uses this distinction
to strengthen his arguments for strong emergence.  We question the correctness of it, however, and thus
the basis for such a classification.  The mechanism for any SSB, whether in QCD due to the scalar
condensate or in cyrstallization due to spontaneous nucleation, originates at the micro, or local, scale,
but the symmetries that are broken are \emph{global}.   This means that the ramifications of the broken
symmetries extend into the macro scale.  How do we know this?  In the case of the crystal we can
definitely observe the long-range order of the crystal's point symmetry, as Ellis correctly points out.
In QCD, on the other hand, we observe pions \emph{everywhere}. Another example is the already mentioned
Higgs field, that is generated in SSB at the electroweak scale and penetrates the {\em whole universe}.

Therefore, there is no distinction between SSB(m) and SSB(M).  There is only just one SSB.  Ellis'
ensuing arguments based off this distinction are thus non sequiturs.

\subsection{We are making progress}
In our original paper we challenged proponents of strong emergence to make a scientific prediction
based off strong emergence that could be tested.  Our goal here was to apply Popper's \emph{Fasifiability}
criterion~\cite{Popper1962-POPCAR}.  Ellis accepted our challenge and answered:

\emph{``Neither LM (Luu \& Mei\ss ner) nor any of their nuclear physics or particle physics
colleagues will be able to derive the experimentally tested properties of superconductivity or
superfluidity in a strictly bottom up way.  In particular, they will be unable to thus derive a successful
theory of high-temperature superconductivity."}

He argues that we will never be able to do this since these phenomena are strong emergent.  Let us be
the first to admit that we (Luu \& Mei\ss ner) have not derived such a theory since the publication of
his challenge, and even with the amount of hubris that we already have, we would never claim that
we ourselves will ever do so.  We stress, however, that our inability to derive such a theory does not provide confirmation of strong emergence.  

But we will point out that some of our colleagues of similar ``ilk''
have made progress in deriving bottom-up theories in areas that Ellis claims belongs to strong emergence.  For example, in~\cite{Kaplan:2019pdd}  a relativistic formulation of the fractional Quantum Hall effect was derived.
Another example is the {\em ab initio} calculation of the so-called Hoyle state in $^{12}$C, which is not only making life on Earth possible, but has also been a barrier
for nuclear theory calculations until 2011~\cite{Epelbaum:2011md}.  Such problems were considered intractable but are now soluble due to advances in high-performance computing and the ingenuity of our colleagues.  
Theoretical physics requires optimism rather than a ``can't do'' attitude.

\section{Conclusions\label{sect:conclusions}}
Conclusions based off misconceptions of EFT have historically lead to erroneous physical
claims  and added  confusion about its ability to explain emergent phenomena as well as its connection
to underlying theories.  Over time many misconceptions have diminished, but unfortunately some still
persist and one must remain vigilant to correct them and the conclusions based off them.  In this
article we corrected the misconceptions of EFT that Ellis uses in his arguments for the case of strong emergence.  

Admittedly, because our imaginations are still too limited (and may be bounded!), we rely heavily on Nature to tell us how to define our physical boundaries that lead to exotic emergent phenomena, especially in cases where the environment is synthetic, like superconductivity in Yttrium-Barium-Copper Oxide.   A proponent of strong emergence would say we ``cheated'', we ``peaked'' at Nature to tell us what to do.  But we make no excuses for this since physics is an \emph{experimental} science after all!

A common argument that Ellis makes related to strong emergent phenomena is that such phenomena would
never be realized from a bottom-up procedure because the environment which enables the phenomena does
not occur naturally in Nature, rather it is synthetic.  He refers to superfluidity and superconductivity
in solid-state physics, which he adamantly claims are strong emergent.  If that is the case, then how
do we classify neutron superfluidity in the crust of neutron stars~\footnote{Here we differentiate
neutron superfluidity that occurs in neutron star crusts where crust nuclei provide a lattice with
phonon interactions, as opposed to the superfluidity that occurs deeper in the star.}, or quark-color
superconductivity predicted to occur deep within dense compact stars~\cite{Haskell:2017lkl,Kobyakov:2013eta}~\footnote{The same could be asked about superconductivity in naturally occurring mercury and lead.}?  The physical mechanisms that underly these phenomena are analogous to those
in solid-state systems, but clearly here the environment is not synthetic.  So are they strong
emergent because they share the same physical mechanism, or are they weak emergent because we predicted the phenomena ourselves?  If we insist that
the solid-state examples are strong emergent, while the others are weak emergent, than doesn't that
imply that the synthetic materials are special?  And by extension, that we humans who created the materials are special?
Such hubris is best avoided, even by us.

\acknowledgements{
TL thanks J.-L. Wynen, C. Hanhart, T. L\"ahde, and M. Hru$\breve{\text{s}}$ka for insightful comments and discussions.  This work was
supported in part by the Deutsche Forschungsgemeinschaft (DFG) through funds provided to
the Sino-German CRC 110 ``Symmetries and the Emergence of Structure in QCD" (Grant No. TRR110), by the
Chinese Academy of Sciences (CAS) through a President's International Fellowship Initiative (PIFI)
(Grant NO. 2018DM0034) and by the VolkswagenStiftung (Grant NO. 93562)}

\end{document}